\documentclass[print, showpacs, twocolumn,superscriptaddress]{revtex4}
\usepackage{graphicx}
\usepackage{amssymb}
\newcommand{\ket}[1]{|#1\rangle}
\newcommand{\bra}[1]{\langle#1|}
\begin{document}

\title{Non-Markovian effect on the quantum discord}
\author{Bo Wang}
\email{bowangphysics@gmail.com} \affiliation{Wuhan Institute of
Physics and Mathematics, Chinese Academy of Sciences, Wuhan 430071,
China} \affiliation{Graduate School of the Chinese Academy of
Sciences, Beijing 100049, China}
\author{Zhen-Yu Xu}
\affiliation{Wuhan Institute of Physics and Mathematics,
Chinese Academy of Sciences, Wuhan 430071, China}
\affiliation{Graduate School of the Chinese Academy of Sciences,
Beijing 100049, China}
\author{Ze-Qian Chen}
\affiliation{Wuhan Institute of Physics and Mathematics,
Chinese Academy of Sciences, Wuhan 430071, China}
\author{Mang Feng}
\email{mangfeng@wipm.ac.cn}
\affiliation{Wuhan Institute of Physics and Mathematics,
Chinese Academy of Sciences, Wuhan 430071, China}

\begin{abstract}

We study the non-Markovian effect on the dynamics of the quantum
discord by exactly solving a model consisting of two independent
qubits subject to two zero-temperature non-Markovian reservoirs,
respectively. Considering the two qubits initially prepared in
Bell-like or extended Werner-like states, we show that there is no
occurrence of the sudden death, but only instantaneous disappearance
of the quantum discord at some time points, in comparison to the
entanglement sudden death in the same range of the parameters of
interest. It implies that the quantum discord is more useful than
the entanglement to describe quantum correlation involved in quantum
systems.
\end{abstract}


\maketitle

Since the proposal by E. Schr\"{o}dinger \cite{entan-Schrodinger},
entanglement has triggered off many famous and imaginative
discussions for our deeper understanding of the quantum world.
Nowadays, entanglement is considered to be not only a vital concept
in physics but also a prime resource for quantum information
processing (QIP) \cite{nielsen}.

Entanglement is something unique without
classical counterpart, which is reflected in the implementation of
quantum teleportation \cite{entan1}, quantum cryptography
\cite{entan2} and universal quantum computing \cite {DiVi}.
However, there seem some exceptions. For example, in Grover search
\cite{Grover0, Knight} and in deterministic quantum computation with
one pure qubit \cite{knill}, entanglement seems unnecessary in the
implementation, although the accomplishment of those quantum tasks
results in something unaccessible in a purely classical way. This
could be understood as that entanglement only represents a special
kind, but not all, of the quantum correlation in those
systems.

The present work focusses on another concept for quantum
correlation, termed quantum discord \cite{zurek,vedral,zurek1},
which was introduced as the difference between two natural quantum
extensions of the classical mutual information. We have noticed some
investigations about the quantum discord to work for quantum
algorithms which showed that quantum discord is more practical than
entanglement to describe quantum correlation
\cite{datta-onequbit,experimental-onequbit}. Besides, quantum
discord could be used to improve the efficiency of the quantum
Carnot engine \cite{Carnot} and to better understand the quantum
phase transition and the process of Grover search \cite{QPT,
Grover}.

It is considered that the quantumness captured by discord is
different from entanglement \cite {zurek,zurek1}. This was also
investigated under the Markovian environment in a recent publication
for dissipative dynamics \cite
{discord-first} that the discord with an asymptotical decrease is
more robust than the entanglement with sudden death under the same
conditions. However, does this characteristic hold under
non-Markovian environmental conditions ?

Since the non-Markovian environment keeps memory, which could
influence the dynamics of the coupled system, it would be of great
interest in an investigation of the dissipative dynamics of the
discord under the non-Markovian environment. To this end, we will
make comparison with `entanglement sudden death' (ESD) \cite {Yu}, a
widely studied terminology reflecting the fragility and complexity
of the entanglement. There have been a lot of investigations on the
ESD \cite{xu11, Bellomo1, xu8, xu7, xu9, xu10, xu12, zeno, Bellomo2,
ESD1}, including some experimental observations for ESD with the
photon pairs and the atomic ensembles \cite{ESD1}. Meanwhile, the
Markovian environment will also be treated here as a comparison. Our
question is whether the quantum discord would present a similar
behavior to the entanglement in the evolution with respect to the
same characteristic parameters of interest.

Our model consists of two independent qubits interacting,
respectively, with their own reservoirs. To be simplified, we assume
the reservoirs to be in zero temperature. As the two qubits are
initially entangled, we will show the dissipative dynamics of the
quantum correlation reflected in the time evolution of the quantum discord in
comparison with the entanglement.

We first present a brief review of the quantum discord. In classical
information theory, the Shannon entropy $H(X)=-\sum_x p_{X=x} \log
p_{X=x}$ is used to measure the uncertainty of a random variable
$X$, where $p_{X=x}$ is the probability with $X$ being $x$.
Similarly, the joint entropy, which measures the total uncertainty
of a pair of random variables $X$ and $Y$, is defined as
$H(X,Y)=-\sum_{x,y} p_{X=x,Y=y} \log p_{X=x,Y=y}$, with
$p_{X=x,Y=y}$ being the probability in the case of $X=x$ and $Y=y$.
As a result, the mutual information for the correlation between two
random variables $X$ and $Y$ is defined as $I(X:Y) = H(X) + H(Y) -
H(X,Y)$, whose quantum version can be written as,
\begin{equation}
{\cal T}(X:Y)=S(\rho_X)+S(\rho_Y)-S(\rho_{XY}), \label{mi2}
\end{equation}
where $S(\rho)=-{\rm Tr}(\rho {\rm log}\rho)$ is the von Neumann entropy
of $\rho$, and $\rho_X(\rho_Y)$ is the reduced
density matrix of $\rho_{XY}$ by tracing out $Y(X)$.

For classical probability distributions, the Bayes rule
$p_{X|Y=y}=p_{X,Y=y}/p_{Y=y}$ leads to an equivalent expression for
the mutual information
\begin{equation}
I(X:Y)= H(X) -  H(X|Y) \label{cmi3},
\end{equation}
where $H(X|Y)=\sum_y p_{Y=y}H(X|Y=y) = -\sum_{x,y} p_{X=x,Y=y} \log
p_{X=x|Y=y}$ is the conditional entropy of the random variables $X$
and $Y$ for the average uncertainty about the value of X given that
the value of Y is known. In order to generalize Eq. (2), we measure
the subsystem $Y$ by a complete set of projectors $\{{\Pi_i}\}$,
corresponding to the outcome $i$, which yields $\rho_{X|i}
={Tr_Y(\Pi_i\rho_{XY}\Pi_i)}/{p_i}$, with $\qquad
p_i=Tr_{XY}(\Pi_i\rho_{XY}\Pi_i)$. So we may define the quantum
conditional entropy $S_{\{\Pi_i\}}(X|Y) = \sum_i p_i S(\rho_{X|i})$.
Following Eq. (\ref{cmi3}), we have the quantum mutual information
alternatively defined by
\begin{equation}
{\cal J}_{\{\Pi_i\}}(X:Y)= S(\rho_X) -S_{\{\Pi_i\}}(X|Y).
 \label{definition2}
\end{equation}
The above quantity strongly depends on the choice of the
measurements $\{\Pi_i\}$. By maximizing ${\cal J}_{\{\Pi_i\}}(X:Y)$
over all ${\{\Pi_i\}}$, we define an independent quantity ${\cal
J}(X:Y)=max_{\{\Pi_i\}}\{S(\rho_X)-\sum_ip_iS(\rho_{X|i})\}\equiv
S(\rho_X) -S(X|Y)$ as a measure of the classical correlation.

Having the two quantum analogs of the classical mutual information
${\cal T}(X:Y)$ and ${\cal J}(X:Y)$, we define the difference
between them as the quantum discord,
\begin{equation}
{\cal D}(X:Y)={\cal T}(X:Y)-{\cal J}(X:Y), \label{definition}
\end{equation}
which is interpreted as a measure of the quantum correlation
\cite{zurek,vedral,zurek1}.

For two noninteracting qubits, i.e., $A$ and $B$ (involving two levels in
each) locally interacting with the reservoirs $R_A$ and $R_B$,
respectively, the Hamiltonian of each subsystem (i.e.,
qubit+reservoir) is given by
\begin{equation}\label{Hamiltonian}
H=\omega_0 \sigma_+\sigma_-+\sum_{k} \omega_k b_k^\dag
b_k+ \sum_{k}(g_{k}b_{k}\sigma_{+} + g_{k}^{*}b_{k}^{\dag}\sigma_{-}),
\end{equation}
where $\omega_0$ denotes the transition frequency of the two-level
system (i.e., the qubit) with $\sigma_ \pm$ the corresponding atomic
raising and lowering operators. The index $k$ labels different field
modes of the reservoir with frequencies $\omega_k$. $b_k^\dag$
($b_k$) is the creation (annihilation) operator of the reservoir
field with $g_k$ the coupling constant to the qubit \cite{Bellomo1,
Bellomo2, open-book}. We assume that the initial state of the qubit
with the zero-temperature reservoir is $|\Psi \left(
0\right)\rangle_{\tilde{S}}=\left( C_{0}(0)|0\rangle
_{S}+C_{1}(0)|1\rangle_{S}\right)|0\rangle_{R_S}$. The amplitudes at
any time can be obtained exactly by,
\begin{eqnarray}
C_{0}(t)=C_{0}(0),  \nonumber\\
\qquad \dot{C}_{1}(t)=-\int_0^tdt^{^{\prime}}F(t-t^{^{
\prime}})C_{1}(t^{^{\prime}}), \label{3}
\end{eqnarray}
where the correlation function $F(t-t^{^{\prime}})=\int d\omega
J(\omega)e^{i(\omega_{0}-\omega)(t-t^{^{\prime}})}$ with $J(\omega)$
the spectral density of the reservoir. The exact form of $C_{1}(t)$
depends on the particular choice of the spectral density of the
reservoir \cite{open-book}. In our model, we use the Lorentzian
spectral distribution
\begin{equation}\label{spectraldensity}
J(\omega)=\frac{1}{2 \pi}\frac{\gamma_0
\lambda^2}{(\omega_0-\omega)^2+\lambda^2},
\end{equation}
where the parameter $\lambda$, defining the spectral width of the
coupling, is connected to the reservoir correlation time $\tau_B$ by
the relation $\tau_B \approx \lambda^{-1}$. For our purpose, we
define another parameter $\gamma_{0}$ regarding the decay of the
atomic excitation in the Markovian limit of the flat spectrum.
The relaxation time scale $\tau_R$ over which the state of the
system changes is then related to $\gamma_0$ by $\tau_R \approx
\gamma_0^{-1}$. The Markovian and the non-Markovian regimes are
distinguished by the relation of the parameters $\gamma_0$ and
$\lambda$. In the Markovian regime there is $\gamma_0<\lambda/2$ or
$\tau_R> 2\tau_B$, and the non-Markovian regime corresponds to
$\gamma_0 > \lambda/2$ or $\tau_R< 2\tau_B$
\cite{Bellomo1,open-book}. In the non-Markovian regime,
straightforward solution of Eq.~(\ref{3}) could yield $C_1(t)\equiv
C_1(0)\chi(t)$ where $\chi(t)=e^{-\lambda t/2}\left[ \cos
\left(dt/2\right)+(\lambda /d)\sin \left(dt/2\right) \right]$ with
$d=\sqrt{|2\gamma_0\lambda-\lambda^2|}$.
In contrast, in the Markovian regime, $C_1(t)$ has the similar form
but with $\cos[\cdot] (\sin[\cdot])$ replaced by $\cosh[\cdot]
(\sinh[\cdot])$. So the dynamics of a qubit $S$ can be represented
by the reduced density matrix
\begin{equation}\label{roA}
\hat{\rho}^S(t)=\left(
\begin{array}{cc}
\rho^S_{11}(0)\chi(t)^2  & \rho^S_{10}(0)\chi(t)\\\\
\rho^S_{01}(0)\chi(t)  & 1-\rho^S_{11}(0)\chi(t)^2 \\
\end{array}\right),
\end{equation}
where $\rho^S_{ij}(0)=C_{i}(0)C^{*}_{j}(0)$, $\chi(t)^2$ has
discrete zeros at $t_n=2\left[n\pi-\arctan (d/\lambda )\right]/d$,
with $n$ being an arbitrary integer. Because the system is composed
of two noninteracting parts evolving independently according to
Eq.~(\ref{Hamiltonian}), we may simply construct the density matrix
$\hat{\rho}$ for the two-qubit system by the reduced single-qubit
density matrices \cite{Bellomo1,Bellomo2}. In the basis $\{\ket{11},
\ket{10}, \ket{01}, \ket{00}\}$, we measure the qubit $B$ from the
matrix $\hat{\rho}(t)$ by projecting on
$\{\cos\theta\ket{1}_B+e^{i\phi}\sin\theta\ket{0}_B,
e^{-i\phi}\sin\theta\ket{1}_B-\cos\theta\ket{0}_B\}$. Then the
quantum discord could be calculated numerically using
Eq.~(\ref{definition}).

The initial states we assume are the extended Werner-like
states (EWL) \cite{werner,Bellomo2}, defined as
\begin{equation}\label{EWL1}
\rho_{EWL}^{\xi}=r\ket{\xi}\bra{\xi}+\frac{1-r}{4}\mathbb{I},
\end{equation}
where $|\xi\rangle = |\Phi\rangle$ or $|\Psi\rangle$ with
$\ket{\Phi}=\alpha\ket{10}+(1-\alpha^2)^{1/2}\ket{01}$
and
$\ket{\Psi}=\alpha\ket{00}+(1-\alpha^2)^{1/2}\ket{11}$.
For $r=0$, the EWL states become totally mixed, while they reduce to
the Bell-like pure state $\ket{\Phi}$ or $\ket{\Psi}$ in the case
of $r=1$.
\begin{figure}
\begin{center}
\includegraphics[width=0.85\linewidth]{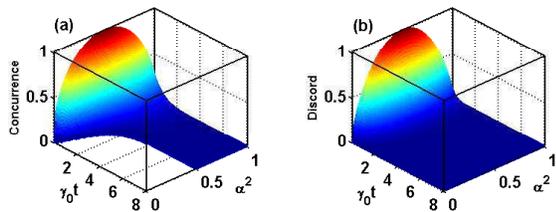}
\caption{(Color online) Variation of the concurrence (a) and the
discord (b) from the initial state $\ket{\Psi}$ with respect to the
dimensionless quantities $\gamma_0 t$ and $\alpha^2$, in the
Markovian case (e.g., $\lambda/\gamma_0=10$).}
\end{center}
\end{figure}

In what follows, we will check if sudden death happens in the
dynamics of the discord under the same condition with respect to the
entanglement. To this end, we employ concurrence representing the
entanglement \cite{Wootters} as a comparison. We get started from
the Markovian regime. In Fig. 1, we plot the dynamics of the
concurrence and the discord for the same Bell-like state
$\ket{\Psi}$ as a function of the dimensionless quantities
$\gamma_0t$ and $\alpha^2$, under the same condition
$\lambda/\gamma_0=10$. There are clearly two ranges for the variance
of the concurrence: For $\alpha^2<1/2$, the ESD occurs, which
denotes the entanglement vanishing abruptly after a finite time; But
for $\alpha^2\geq 1/2$, the concurrence vanishes in an asymptotical
way. In contrast, the discord vanishes only asymptotically with the
variance of those characteristic parameters.
\begin{figure}
\begin{center}
\includegraphics[width=0.85\linewidth]{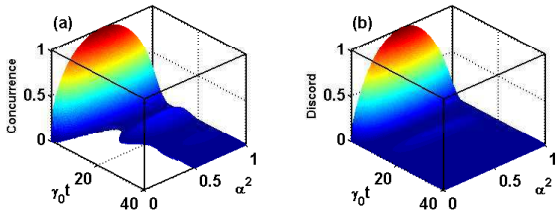}
\caption{(Color online) Variation of the concurrence (a) and the
discord (b) from the initial state $\ket{\Psi}$ with respect to the
dimensionless quantities $\gamma_0 t$ and $\alpha^2$ in the
non-Markovian case (e.g., $\lambda/\gamma_{0}=0.1$).}
\end{center}
\end{figure}

Fig. 2 presents the case for the non-Markovian regime (i.e.,
$\lambda/\gamma_0=0.1$) from the initial Bell-like state
$\ket{\Psi}$. It can be seen that the discord of $\ket{\Psi}$
periodically vanishes in accordance with the zero points of the
function $\chi(t)^2$ following the asymptotical damping. On the
contrary, there is ESD for the concurrence under the same condition
in some ranges of the parameter $\alpha^2$. A clearer difference
between the behaviors of the discord and the concurrence could be
found in Fig. 3(a).
\begin{figure}
\begin{center}
\includegraphics[width=0.9\linewidth]{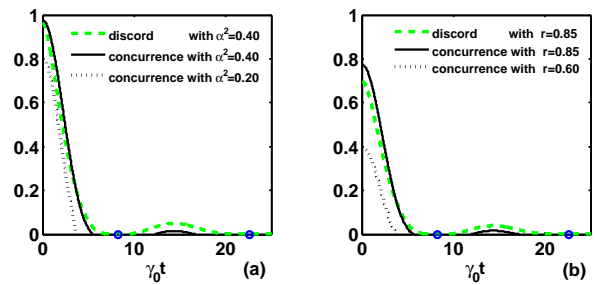}
\caption{(Color online) The comparison of the evolution of the
concurrence with the discord in the non-Markovian case (e.g.,
$\lambda/\gamma_0=0.1$): (a) For the initial state $\ket{\Psi}$ and
(b) for the initial mixed state $\rho_{EWL}^{\Psi}$. The green
dashed line denotes the discord. The black (dotted and solid) curves
correspond, respectively, to the concurrence with different
parameters $\alpha^2$ or $r$. $\circ$ denotes the zero points of the
discord.}
\end{center}
\end{figure}

To see what happens in the mixed state, we have made a study in Fig.
4 for two initial EWL states $\rho_{EWL}^{\Phi}$ and
$\rho_{EWL}^{\Psi}$ with $\alpha=1/\sqrt{2}$ in the non-Markovian
regime. We have found that the concurrence exists at the beginning
of the evolution (i.e., $t=0$) only for $r> 1/3$, while the discord
is always positive for $r>0$. The common feature of the dynamics for
both the concurrence and the discord is the decrease with the
decrease of the purity $r$, implying that the mixedness of the
initial state affects both of them. The difference between them is
similar to the case of the pure state, as shown more clearly in Fig.
3(b).
\begin{figure}
\begin{center}
\includegraphics[height=6cm,width=0.9\linewidth]{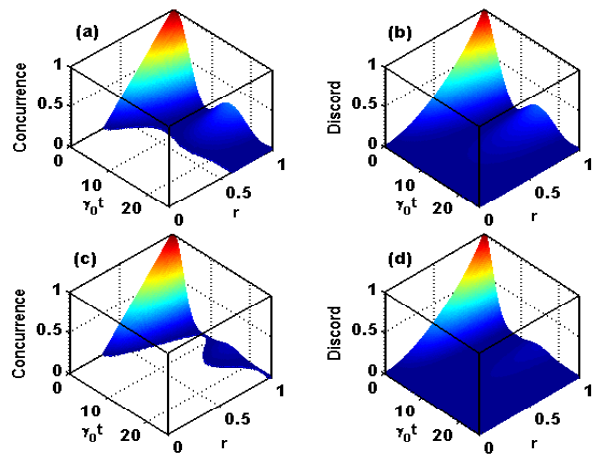}
\caption{(Color online) Variation in the non-Markovian case with
respect to the dimensionless quantities $\gamma_{0}t$ and $r$, where
$\lambda/\gamma_0=0.1$ and $\alpha^2=1/2$, (a) and (b) are for the
concurrence and the discord, respectively, from the initial states
$\rho_{EWL}^{\Phi}$; (c) and (d) correspond to the same variables as
in (a) and (b), but from the initial states $\rho_{EWL}^{\Psi}$.}
\end{center}
\end{figure}

How much the non-Markovian environment influences the discord? Since
$\lambda/\gamma_0$ characterizes the degree of the non-Markovian
effects, we have studied the dynamics of the discord in Fig. 5
starting from the EWL states $\rho_{EWL}^{\Psi}$ with $r=1$ and
$\alpha=1/\sqrt{3}$. We can observe that the revival amplitude of
the discord increases with the decrease of the $\lambda/\gamma_0$
corresponding to the enhancement of the non-Markovian effects.

Our results for the non-Markovian regime have shown something very
different from the Markovian regime \cite {discord-first}: The
discord disappears instantaneously at some time points following the
asymptotical dissipation. In some sense, this could also be called
sudden death and sudden revival of the discord, but we prefer to
call it the instantaneous disappearance. Although the mechanism behind the
instantaneous disappearance needs further clarification, some
points seem clear to us: The revival is due to the memory effect of
the non-Markovian reservoir. The change of $\lambda$ in Fig. 5
implies that the stronger memory effect of the reservoir yields the
larger revival amplitude of the discord. Besides, the zero discord
in our model happens in the case of the evolution to the component
state $|00\rangle$, $|01\rangle$, $|10\rangle$, $|11\rangle$ or the
maximal mixed state $\mathbb{I}$. In particular, for some values of
the parameters, such as $\alpha=0$ or 1 in Figs. 1 and 2, and $r=0$
in Fig. 4, the discord remains zero in the evolution. In contrast,
for other case, the discord only falls asymptotically, with some
fluctuations reaching zero instantaneously.

\begin{figure}
\begin{center}
\includegraphics[width=0.85\linewidth]{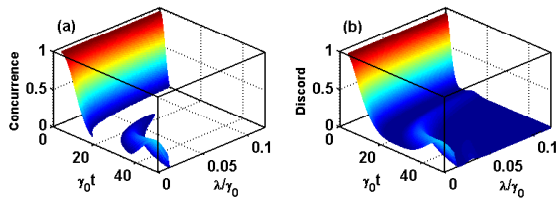}
\caption{(Color online) Variation of the concurrence (a) and the
discord (b) from the initial states $\rho_{EWL}^{\Psi}$ with respect
to the dimensionless quantities $\gamma_0 t$ and $\lambda/\gamma_0$
in the non-Markovian case, where the parameters $r=1$ and $\alpha^2=1/3$
are set in the calculation.}
\end{center}
\end{figure}

The Lorentzian spectral distribution in Eq. (7) had been widely
employed in quantum optics \cite{open-book}, and was recently used
in QIP studies \cite {Bellomo1}. Since strong coupling between
matter and light has been available experimentally in some systems
\cite{dublin}, dynamics on short-time scale could in principle be
observed and even be manipulated. As a result, non-Markovian
effect would become more and more important in the exploration of
QIP under real experimental environment. Moreover, our studies for
both Markovian and non-Markovian reservoirs have clearly
demonstrated that the discord is more robust in dissipative
evolution with respect to entanglement and more suitable to describe
the quantum correlation involved in the system. As a result, it is
understandable why some quantum algorithms could work well in the
absence of entanglement. The discord should be larger than zero in
the implementation of those algorithms in the case that the
entanglement approaches zero. A deeper study in this respect is
expected.

In conclusion, we have investigated the dynamics of the discord
using an exactly solvable model where each qubit independently
interacts with its own zero-temperature reservoir. We have compared
the discord with the concurrence using the same initial states and
the same reservoir conditions. We have also discussed the different
effects from the Markovian and non-Markovian reservoirs. Further work
would be to explore the dynamics of the discord subject to some
specific decoherence mechanism, and to quantitatively evaluate the
necessary discord required in implementing different quantum
algorithms. We will also consider the situation at finite
temperature. In comparison with some recent work on non-Markovian
effect on the dissipation of the system in a microscopic way \cite
{zhang}, our present work from the phenomenological viewpoint, might
be more practical to explain some experimental observations of the
dissipation of  quantum correlation subject to a realistic
environment.

This work is supported by NNSFC under No. 10774163, No. 10774042 and
No. 10775175.

Note added: The work was submitted to Physical Review A in September.
But we have just been aware of a similar work in arXiv:0911.1096v2.

\end{document}